\documentclass[showpacs,twocolumn,prl,floatfix,amssymb,aps,superscriptaddress]{revtex4}
\usepackage{graphicx}

\newcommand{\half}{\hbox{$\frac{1}{2}$}}

\begin{document}

\hsize\textwidth\columnwidth\hsize\csname@twocolumnfalse\endcsname

\title{Insulator to superfluid transition in coupled photonic cavities in two dimensions}

\author{Jize Zhao}
\affiliation{Institute for Solid State Physics, University of Tokyo, Kashiwa, Chiba 277-8581, Japan}

\author{Anders W. Sandvik}
\affiliation{Institute for Solid State Physics, University of Tokyo, Kashiwa, Chiba 277-8581, Japan}
\affiliation{Department of Physics, Boston University, Boston, Massachusetts 02215, USA}

\author{Kazuo Ueda}
\affiliation{Institute for Solid State Physics, University of Tokyo, Kashiwa, Chiba 277-8581, Japan}

\date{\today}

\begin{abstract}
A two-dimensional square lattice of coupled photonic cavities is  systematically investigated using quantum Monte Carlo simulations. The ground state phase 
diagram contains insulating phases with integer polariton densities surrounded by a superfluid phase. Finite-size scaling of the superfluid stiffness is 
used to extract the phase boundaries. The critical behavior is that of the generic, density-driven insulator-superfluid transition with dynamic exponent $z=2$, 
with no special multicritical points with $z=1$ at the tips of the insulating-phase lobes (in contrast to the Bose-Hubbard model). This demonstrates 
a limitation of the description of polaritons as structureless bosons.
\end{abstract}

\maketitle

One of the currently most active and intriguing areas of condensed matter physics is the search for ``exotic'' quantum phases and 
related quantum phase transitions \cite{SACH1} in strongly correlated systems.  However, the control of the relevant system parameters, which 
correspond to various coupling constants of theoretical model systems, remains restricted in chemically synthesized materials. This 
has stimulated investigations of artificially engineered nano-structures, in which a wider range of model behavior can be realized. 
Of these systems, coupled Josephson junctions and ultracold atoms in optical lattice are perhaps the most prominent so far \cite{ZANT1,JAKS1,LEWE1}. 
The latter case is, among other things, ideally suited for implementing the Bose--Hubbard model \cite{FISH1} throughout its parameter space. 

Recently, another engineered  structure was proposed which shows great promise for investigating novel quantum states of matter; arrays of optical 
cavities in each of which there is one or several atoms interacting with photons according to the famous Jaynes-Cummings interaction \cite{HART1,GREE1,AMGE1}. 
The cavities are coupled through photon tunneling (hopping) between adjacent cavities. A particular advantage of this structure over atoms in optical lattices is that 
the distance between the cavities is comparably large \cite{HART1}, which allows for the manipulation and measurement of individual cavities experimentally. 
In these systems, the number of photons is not conserved, but the number of polaritons---quasiparticles constructed as combinations of photons and atomic 
excitations---are conserved and have been argued to be analogous to the particles of the Bose-Hubbard model. Theoretically, it has been well established that there exists a 
polariton localization--delocalization transition \cite{HART1,GREE1,AMGE1,HUO1,ROSS1, MARK1}, in analogy with the insulator--superfluid transition in the 
Bose-Hubbard model \cite{FISH1}. Extending the analogy, a system with two species of photons has also been investigated and related to the 
two-component Bose--Hubbard model \cite{HART2}, and a super-counter-fluidity phase has been predicted for circularly polarized photonic cavities \cite{JI1}. While
experimental work on coupled optical cavities has not yet reached the stage where these states could be explored, Bose-Einstein condensation of polaritons has been 
demonstrated in single two-dimensional cavities with a large number of embedded quantum wells \cite{KASP1}.

In this paper we will focus on the simplest version of optical cavities on the two-dimensional (2D) square lattice; in each cavity there is a single atom with two energy 
levels and the photons are polarized so that we need to consider only one-component photons. This system is described by the Hamiltonian
\begin{equation}
\mathcal{H} =-t\sum_{\langle{ij}\rangle}(a^{\dagger}_{i}a_{j}+h.c.) + \sum_{i}h_i,
\label{HHOP}
\end{equation}
where $a_i$ and $a^{\dagger}_{i}$ are photon annihilation and creation operators, and the single-site terms $h_i$ are given by
\begin{eqnarray}
&&h_{i}=\epsilon(S^{z}_{i}+\half)+\omega a^{\dagger}_{i}a_{i} - \mu(S^{z}_{i}+\half+a^{\dagger}_{i}a_{i}) \nonumber \\
&& ~~~~~~~~ +g(S^{+}_{i}a_{i}+S^{-}_{i}a^{\dagger}).
\label{HSITE}
\end{eqnarray}
Here a two-level atom is represented by a spin-$1/2$, with $S^{\pm}_{i}$ being the corresponding raising and lowing operators. The atomic excitation energy and the 
resonance frequency of the cavity are denoted by $\epsilon$ and $\omega$, respectively, and $g$ is the coupling constant between photons and atoms. The excitations of 
isolated cavities ($t=0$) are not photons but entangled photon--atom states---the polaritons. Their total number operator is given by
\begin{equation}
\mathcal{N}=\sum_{i}n^{\rm pol}_{i}=\sum_{i}(a^{\dagger}_{i}a_{i}+S^{z}_i+\half),
\end{equation}
which commutes with Hamiltonian, $[\mathcal{N},\mathcal{H}]=0$, for all $t$. Thus $\mu$ in (\ref{HSITE}) is the chemical potential for polaritons.

Exact numerical calculation with the density-matrix-renormalization-group (DMRG) method for the 1D case has confirmed a ground state phase diagram with insulating and superfluid 
phases \cite{ROSS1}. A mean-field theory showed a similar behavior in 2D \cite{GREE1}. In both 1D and 2D, the insulating phases have integer polariton density and form a series 
of lobes in the $(t,\mu)$ plane, in qualitative agreement with the Bose-Hubbard model. Very recently, Aichhorn {\it et al.} presented 1D and 2D phase diagrams, as well as 
single-particle spectra, obtained with a systematically controllable variational cluster approach \cite{MARK1}. Their 1D results are in good agreement with the DMRG calculations. 

The previous numerical calculations have convincingly demonstrated the existence of the insulating and superfluid phases and located their phase boundaries, but the exact nature 
of the phase transition has not yet been determined. A major difference between the cavity model (\ref{HHOP}) and the Bose-Hubbard model is that there is no direct on-site 
repulsion between photons or polaritons, which plays a key role \cite{FISH1} in the insulating phase of the Bose-Hubbard model (although there is an effective repulsion between photons 
mediated by the spins \cite{BIRN1}). Perhaps more importantly, the polaritons are complex quantum objects different from pure bosons. It is therefore not {\it a priori} clear that 
the phase transition between the insulating and superfluid phases should be in the same universality class as in the Bose-Hubbard model. 

Here we investigate the quantum phase transition numerically, 
using approximate-free quantum Monte Carlo simulations (stochastic series expansion with loop updates \cite{SAND1}). We use finite-size scaling to extract the phase boundaries, as 
well as exponents governing the quantum-critical scaling behavior, and conclude that there is indeed a qualitative difference between the two models. In the Bose-Hubbard model there are special 
multi-critical points at the tips of the insulating lobes, where the transition occurs at constant density and the critical behavior is non-generic, with 3D classical XY exponents---most
notably the dynamic exponent $z=1$ at the tips. In contrast, the transition in the optical cavity system is always of the generic, density-driven mean-field kind, with dynamic exponent 
$z=2$. Another interesting aspect of the phase diagram is the shape of the lobe tips, which are cusped in 1D \cite{ROSS1}. The calculations of Aichhorn {\it et al.}~also indicate cusps 
in 2D \cite{MARK1}, although much less pronounced than in 1D, in contrast to the smooth tips in the 2D Bose-Hubbard model \cite{FISH1,CAPO1}. However, our calculations show that the
lobes are smooth.

We consider two kinds of phase diagrams: (i) at chemical potential $\mu=0$ in the $(\Delta,{t})$ plane, where $\Delta=\epsilon-\omega$ is the detuning parameter, and (ii) in the 
$(\mu,{t})$ plane with the detuning $\Delta=0$ (as studied also in \cite{MARK1}). For simplicity we set $g=\epsilon=1$. When the hopping $t=0$, the Hamiltonian can be readily 
diagonalized and we can extract the number of polaritons $n^{\rm pol}>0$ as a function of the other parameters. For finite $t$, only the total number of polaritons $\mathcal{N}$ 
is conserved, and the average number $\langle n^{\rm pol} \rangle$ is in general non-integer. However, it is integer in the insulating phases at temperature $T=0$. In our QMC simulations 
we have to impose a maximum polariton number $n^{\rm pol}_{\rm max}$ per site. The cut-off error is systematically checked by comparing results for $n^{\rm pol}_{\rm max} = 3,4,5,6$. 
We consider only the insulating phases with $n^{\rm pol}=1,2$ and find no detectable effects of the cut-off once 
$n^{\rm pol}_{\rm max} \ge 4$. 

\begin{figure}
\includegraphics[width=7cm, clip]{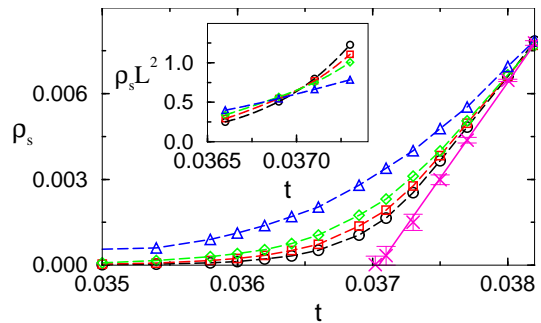}
\null\vskip-4mm
\caption{(Color online) Superfluid density versus hopping for $L=14~(\triangle)$, $18~(\Diamond)$, $20~(\square)$, and $22~(\bigcirc)$, at $(\mu,\Delta)=(0,0.3)$.
Error bars are at most comparable with the symbol size. The data extrapolated to $L=\infty$, using an $L^{-2}$ finite-size dependence, are shown ($\times$) along with a 
fitted line. The inset shows the finite-size behavior corresponding to $z=2$.}
\label{FIG1}
\vskip-3mm
\end{figure}

Using the finite-size-scaling hypothesis for the superfluid density as a function of $L$, $\beta=1/T$, and $\delta=t-t_c$ \cite{FISH1},
\begin{equation}
\rho_{s}=L^{2-d-z}\tilde{\rho}(\delta L^{1/\nu}, \beta/L^z),
\label{rhoscale}
\end{equation}
we can determine a critical insulator--superfluid transition point accurately by measuring $\rho_s$ (in the standard way through the photon winding number fluctuations) near the critical point 
for different lattice sizes $L\times L$ and hopping $t$. If we fix $\beta/L^{z}=\alpha$, then the scaling function $\tilde\rho()$ is a function of the single parameter $\delta L^{1/\nu}$. 
At the critical point $\delta$ is zero and $\rho_{s}L^{d+z-2}$ is then independent of lattice size, which implies that calculated $\rho_{s}L^{d+z-2}$ curves versus $t$ for different $L$ 
should intersect at the critical point. In our analysis we test $z=2$, the generic dynamical exponent of Bose-Hubbard model, and, as already mentioned, find this to hold on all points of 
the phase boundaries. We use two different aspect ratios, $\alpha=1/3$ and $1/2$, for added confidence in the results.

We first show in Fig.~\ref{FIG1} examples of the behavior of the superfluid density close the phase transition. Results for the superfluid density in the range $t=0.0350-0.0382$ at 
fixed $(\mu,\Delta)=(0.0,0.3)$ are graphed for lattice sizes $L=14,18,20$, and $22$.  Extrapolating the results to $L=\infty$ (using $\propto L^{-2}$ finite-size corrections, which our 
data exhibit), we observe the superfluid density in the thermodynamic limit becoming non-zero for $t>t_c=0.03702(5)$. Above this critical point the superfluid density grows linearly with 
$\delta$, which if $z=2$ is consistent with the generalized Josephson relation $\rho_s\sim\delta^{\nu(d+z-2)}$ in combination with the hyperscaling relation $\nu=1/z$ \cite{FISH1}. 

\begin{figure}
\includegraphics[width=7cm,angle=0]{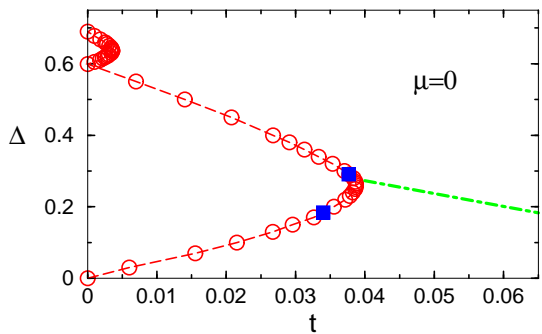}
\null\vskip-4mm
\caption{(Color online) Phase boundary between the insulating and superfluid ground states in the $(t,\Delta)$ plane for polariton chemical potential $\mu=0$. The squares 
show two points determined by finite-size scaling at fixed $t$, varying $\Delta$. The other points were extracted by varying $t$ at fixed $\Delta$. The dot-dashed line shows the 
$\langle n^{\rm pol}\rangle = 1$ polariton density contour.} 
\label{FIG2}
\vskip-3mm
\end{figure}

The finite-size scaling of the superfluid density at criticality, $\rho_s \propto L^{-z}$ with $z=2$, is shown in the inset of Fig.~\ref{FIG1}. The critical hopping $t_c$ is readily
determined from the crossing point (more accurately than using the extrapolated $\rho_s$ above). We use this procedure to determine the full phase boundaries of the first two 
insulating lobes in the $(t,\Delta)$ plane at $\mu=0$, shown in Fig.~\ref{FIG2}. For comparison we also determine some critical points by scanning along fixed-$t$ lines across 
the first insulating lobe (instead of scanning $t$ at fixed $\Delta$). As seen in the figure, the phase boundaries obtained by the two 
methods are in excellent agreement. The lobe tips are located at $(t,\Delta)=(0.03872(5),0.260(5))$ for $n^{pol}=1$ and $(t,\Delta)=(0.00358(4),0.636(3))$ for $n^{pol}=2$. 
The $t$ at the first lobe tip is about $11$ times larger than at the second one, and, thus, the lobes shrink more quickly than the $1/n$ behavior of the Bose-Hubbard model. This 
had already been observed in the 1D case \cite{ROSS1,MARK1}. Fig.~\ref{FIG2} also shows the constant-density $\langle n^{\rm pol}\rangle=1$ contour inside the superfluid phase. 
Here this curve does not connect to the tip of the lobe, as it would have to do (as a requirement for positive compressibility \cite{FISH1}) in a $(t,\mu)$ phase diagram---we will discuss such 
a case below. In the Bose-Hubbard model, the universality class of the transition is different exactly at the tip of the lobe, because density fluctuations are suppressed there due to an 
emergent particle-hole symmetry. This special transition therefore belongs to the three-dimensional XY universality class, whence $z=1$. However, here we find $z=2$ throughout, as we will 
comment on more extensively further below.

\begin{figure}
\includegraphics[width=7cm,angle=0]{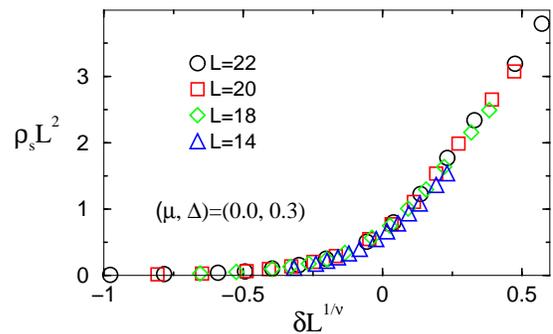}
\null\vskip-4mm
\caption{(Color online) Finite-size data collapse of $\rho_{s}L^{z}$ as a function of $(t-t_c)L^{1/\nu}$, with $z=2$ and $\nu=1/2$, showing consistency
with with the hyperscaling relation $\nu=1/z$.}
\vskip-3mm
\label{FIG3}
\end{figure}

Hyperscaling requires that the correlation length exponent $\nu=1/z$, and thus we should have $\nu=1/2$. The onset superfluid density shown in Fig.~\ref{FIG1} already demonstrated
consistency with $z=2$ and $\nu=1/2$. We can test these exponents to higher precision by a standard data collapse procedure, graphing $\rho_s L^2$ versus $\delta L^{1/\nu}$ for 
several system sizes. A typical example demonstrating the expected scaling is shown in Fig.~\ref{FIG3}. 

\begin{figure}
\includegraphics[width=7.1cm]{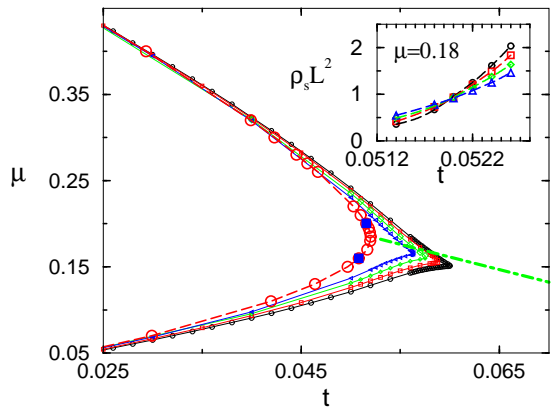}
\null\vskip-4mm
\caption{(Color online) First-lobe boundary in the $(t,\mu)$ plane for $\Delta=0$. The open circles show results of finite-size scaling using scans of $t$ at fixed $\mu$, while the 
two solid squares were obtained at fixed $t$, scanning $\mu$. The other curves are from the variational cluster calculations in Ref.~\cite{MARK1}, with increasing cluster size corresponding 
to the curves right to left (see \cite{MARK1} for details). The dot-dashed line touching the lobe tip is the $\langle n^{\rm pol}\rangle=1$ contour. The inset demonstrates scaling of the 
superfluid density with $z=2$ close to the tip of the lobe, using data for $L=16,18,20,22$.}
\label{FIG4}
\vskip-3mm
\end{figure}

Aichhorn {\it et al.}~considered a slightly different phase diagram from the one we have focused on above. They calculated the phase boundary in the $({t},\mu)$ plane using
a variational cluster method \cite{MARK1}, which should become exact in the limit of large embedded clusters. In order to compare one-to-one with their results, we also extracted 
the boundary of the first insulating lobe in the same parameter space, using the same finite-size scaling methods discussed above. Our results close to the first lobe tip are 
shown in Fig.~\ref{FIG4}, along with the data of Ref.~\cite{MARK1} for different cluster sizes. At small hoppings, the phase boundaries obtained with the two methods agree 
very well, and there is very little dependence on the cluster size in the variational calculation. However, close to the lobe tip there are notable differences. Here the dependence 
on the cluster size in the variational calculation is much more pronounced, but our phase boundary does appear to be roughly consistent with the behavior for increasing cluster size. 
However, there is a qualitative difference in that we obtain a smooth lobe tip, instead of the cusps that appear persistently for all cluster sizes in the variational calculation. It 
is quite possible that the cusps scale away with increasing cluster size, but carrying out this finite-size scaling would be difficult in practice, due to the limited range of available 
cluster sizes. We have checked our calculations at the lobe tip extensively, carrying out the finite-sizse scalings using horizontal as well as vertical parameter scans, looking for 
possible drifts in the $\rho_s L^2$ crossing points, and checking the dependence on the cut-off $n^{\rm pol}_{\rm max}$. We do not find any indications of a cusp and conclude that the cusps are
artifacts of small cluster sizes in the variational calculation. The figure also shows the $\langle n^{\rm pol}\rangle=1$ density contour in the superfluid phase, which connects to the 
tip of the lobe, as it should \cite{FISH1}.

The inset of Fig.~\ref{FIG4} shows the finite-size scaling of the superfluid density very close to the lobe tip. The location of the tip is determined to $t=0.05201(5)$,
$\mu=0.185(5)$. Also here the scaling with $z=2$ is excellent---there are no indications of drifts of the crossing points with increasing lattice size. Thus, we conclude that
there is no special $z=1$ multi-critical point. Although this different universality, if present, should only exist at a single point, 
it should be reflected as a $z=1 \to z=2$ finite-size cross-over in an extended region surrounding that point. In the Bose-Hubbard model $z=1$ scaling close to the lobe tip can 
be seen very clearly for the same range of system sizes studied here \cite{CAPO1} (larger sizes were studied in this paper, but $z=1$ scaling is seen already for $L=10$). 
The fact that we observe essentially perfect scaling with $z=2$ everywhere is therefore very strong evidence in favor of a generic, mean-field density-driven transition also at 
the tip of the lobes. 

In summary, we have studied the quantum phase transition separating a localized (insulating) and delocalized (superfluid) state in a 2D system of coupled optical cavities. Our main 
result is that this transition always has dynamic exponent $z=2$, corresponding to the generic mean-field insulator-superfluid transition of interacting bosons. There are 
no special multi-critical points with $z=1$, which do exist in the Bose-Hubbard model at the tips of the insulating-phase lobes. 
The multi-criticality, of the XY universality class, in the Bose-Hubbard model is a result of suppression of density fluctuations upon approaching the lobe-tips \cite{FISH1}. In 
the coupled optical cavities the polariton density fluctuations should also freeze out at the lobe tips, but the photon number is not conserved. The superfluid stiffness that we
have studied here is defined soley in terms of the photons (calculated using their winding number fluctuations), and thus it appears that the superfluid is a state essentially 
describable in terms of the photons only (with the atomic excitations acting effectively as a fluctuating background potential). Although the effective particles in the insulating 
state are clearly polaritons, it is plausible that the phase transition can be described solely in terms of the photons, because although their number fluctuates in the insulating 
phase the essence of the transition is just that they become localized. Our results that $z=2$ everywhere supports this notion of a generic photon delocalized--localized transition. 
On a microscopic level, the localization still of course involves the formation of polaritons, but the details of the localization mechanism are likely irrelevant (as long as the 
photon density fluctuations are not suppressed, which would lead to XY criticality). There has been recent work based on a very small number (2-6) cavities to elucidate the nature of 
the effective particles in different parameter regimes \cite{IRI1,MAK1} of the coupled optical cavities. Two types of superfluids, photonic and polaritonic, were discussed\cite{IRI1}. However, 
it remains to be clarified whether there really is any clear distinction between the two, in particular as regards the phase transition. In principle the method we have used here 
could also be used to probe the mechanism of localization on a more microscopic level, as a complement to studying the resulting macroscopic critical behavior.

J.~Z. is supported by the Japan Society for the Promotion of Science (P07036). A.~W.~S is supported by the NSF under grant No.~DMR-0513930 and also would like to thank 
the ISSP for support and hospitality during an extended visit. The computation in this work has been done using the facilities of the Supercomputer Center, Institute for 
Solid State Physics, University of Tokyo.

\null
\vfill

\end{document}